\documentclass[ reprint,
amsmath,amssymb,
aps,
]{revtex4-2}

\usepackage{epsf,epsfig}
\usepackage{amssymb,amsmath,amsfonts}
\usepackage{color}
\usepackage{braket}
\usepackage{subcaption}
\usepackage[dvipsnames, svgnames,  x11names ]{xcolor}
\usepackage[T1]{fontenc} % if needed
\usepackage[colorlinks]{hyperref}
\hypersetup{
   citecolor = {blue},
   linkcolor = {RubineRed},
}

\usepackage{physics}
\usepackage{tikz}
\usetikzlibrary{arrows,snakes,backgrounds}

\usepackage[english]{babel}

\usepackage{graphicx}% Include figure files
\definecolor{NordCyan}{HTML}{8FBCBB}          % nord7
\definecolor{NordBrightCyan}{HTML}{88C0D0}    % nord8
\definecolor{NordBlue}{HTML}{81A1C1}          % nord9
\definecolor{NordBrightBlue}{HTML}{5E81AC}    % nord10
\definecolor{NordRed}{HTML}{BF616A}           % nord11
\definecolor{NordOrange}{HTML}{D08770}        % nord12
\definecolor{NordYellow}{HTML}{EBCB8B}        % nord13
\definecolor{NordGreen}{HTML}{A3BE8C}         % nord14
\definecolor{NordMagenta}{HTML}{B48EAD}       % nord15

\begin{document}

\title{Entanglement Renormalization of the class of  Continuous Matrix Product States}

%\preprint{....}

\author{Niloofar Vardian}
\email{nvardian@sissa.it}
\affiliation{SISSA, International School for Advanced Studies, via Bonomea 265, 34136 Trieste, Italy\\
INFN, Sezione di Trieste, via Valerio 2, 34127 Trieste, Italy}

\begin{abstract}
\noindent
Continuous tensor network gives a variational ansatz for the ground state of the quantum field theories (QFTs). The notable examples are the continuous matrix product state (cMPS) and the continuous multiscale entanglement renormalization ansatz (cMERA). While cMPS is just adapted to the non-relativistic QFTs, only the Gaussian cMERA  is well-understood which we can not use to approximate the ground state of the interacting relativistic QFTs. But instead, cMERA also corresponds  to a real-space  renormalization group  flow  in the context of the wave functions. In this letter, we investigate the backward Gaussian cMERA renormalization group  flow of the class of cMPS by putting the standard cMPS at the IR scale. At the UV scale, for the bosonic systems in the thermodynamic limit, we achieve the variational class of states that has been proposed recently as the relativistic cMPS (RCMPS)  is adapted to the relativistic QFTs without requiring to introduce of any additional IR or UV cut-off. We also extend the RCMPS to fermionic systems and theories on a finite circle. 

\end{abstract}
\maketitle

Tensor Network states are the entanglement-based ansatz that has arisen in recent years based on the renormalization group (RG) ideas and  later on developed using tools and concepts from quantum information theory. 
The main examples include 
\emph{matrix product states}
(MPS) \cite{Fannes:1990ur},
\emph{projected entangled-pair states} (PEPS) 
\cite{Verstraete:2004cf}, and \emph{multiscale entanglement renormalization ansatz}  (MERA) \cite{Vidal:2007hda}. 
By construction, they obey the entropy/area law \cite{verstraete2006matrix,osborne2006efficient, hastings2007area,eisert2010colloquium} and are able to encode both global and local symmetries \cite{sanz2009matrix,haegeman2015gauging, zohar2016building, kull2017classification, molnar}. Therefore, they provide an efficient class of symmetric variational ansatz to approximate the ground state of the local Hamiltonian.
In general,  the understanding of the low-energy behavior of many-body quantum systems is one of the major challenges of modern physics, both in high energy and condensed matter physics. There are plenty of methods based on RG introduced to tackle this problem. 
To study the weakly coupled system, one can use the momentum space RG \cite{dyson1949s, gell1954quantum,wilson1971renormalization, wilson1971renormalization2, wilson1975renormalization}. But instead, in the case of the strongly interacting systems where the perturbation theory fails, this question is usually addressed by real-space RG methods.

In the case of the many-body system on the lattice, Kadanoff's spin-blocking idea \cite{kadanoff1966physics} was replaced by Wilson's real space RG \cite{wilson1975renormalization} which is improved later by White's 
\emph{density matrix renormalization group} (DMRG) \cite{white1992density, white1993density}. 
This technique is extraordinarily powerful in the studying of quantum systems on the 1-D lattice.
It has been generalized as
\emph{ tensor renormalization group}
 (TRG) by Levin and Nave \cite{levin2007tensor} to study the Euclidean path integral of 1-D quantum systems or the 2-D classical lattice models.
Although, both the DMRG and TRG are very successful, they provide a coarse-grained system that still contains irrelevant microscopic information which implies the breakdown of both methods at criticality \cite{levin2007tensor}, and the resulting RG flow has the wrong structure of noncritical fixed points \cite{gu2009tensor}.
In the context of wave functions, this problem was resolved with the introduction of 
\emph{entanglement renormalization}
(ER) by Vidal \cite{Vidal:2007hda}. 
A key aspect of ER is the ability to remove the short-range entanglement at each coarse-graining step by introducing a disentangler operator. 
This leads to restoration of scale-invariant at criticality and results in a proper RG flow with the correct structure of fixed points both at criticality and off criticality. 
More recently, this technique has been adapted to tackle the same problem in TRG in the context of the Euclidean path integral of the quantum many-body systems and the partition function of a classical statistical system by removing short-range correlations this time from the partition function, known as 
\emph{tensor network renormalization }
(TNR) \cite{evenbly2015tensor1}. 
ER and TNR  represent a powerful alternative to Wilsonian real-space RG methods  in the context of the wave function and partition function respectively.

Beginning with the DMRG, it  has been shown that this technique can be understood as a variational method within the class of MPS \cite{ostlund1995thermodynamic}. In addition, it justifies the point that the DMRG is powerful just in one spatial dimension because of the area law.
% And, by using the relation between the tensor network states and real space RG, the extension of DMRG can be done by finding the generalization of the MPS.
More generally, any variational class corresponds to a RG scheme. As another important example, the ER is naturally associated with the class of  MERA \cite{Vidal:2007hda}. 

% holography MERA \cite{swingle2012entanglement, evenbly2011tensor, beny2013causal, molinajfep, matsueda2013tensor, nozaki2012holographic, hartman2013time, czech2016tensor}

% holography cMERA \cite{miyaji2015continuous, miyaji2017path, mollabashi2014holographic, caputa2017anti, molina2015information, molina2016entanglement}

Tensor Network formalism can be also applied to study the low-energy limit of quantum field theories (QFTs), after an appropriate discretization of the theory on the lattice \cite{milsted2013matrix, banuls2013mass, banuls2016chiral, banuls2017density, kadoh2019tensor, delcamp2020computing}.
However, the symmetries of spacetime in this way will be destroyed. 
Thus, it would be desirable to work directly in the continuum which can provide a powerful non-perturbative approach for studying the strongly interacting QFTs.
In the last decade, the generalization from lattice to continuum has been done for some classes of tensor network states \cite{verstraete2010continuous, haegeman2011, Hu:2018hyd, tilloy2019continuous, Shachar:2021vbu}. In particular, the continuous version of MPS and MERA, known as cMPS \cite{verstraete2010continuous} and cMERA \cite{haegeman2011}. 
To date, only the Gaussian cMERA is well-understood, which limits the interest of cMERA to use  as a variational ansatz to study the strongly coupled QFTs. Instead, cMERA 
%is actively investigated in the context of holography
 has already attracted considerable
attention in the context of holography
\cite{swingle2012entanglement, evenbly2011tensor, beny2013causal, molinajfep, matsueda2013tensor, nozaki2012holographic, hartman2013time, czech2016tensor, miyaji2015continuous, miyaji2017path, mollabashi2014holographic, caputa2017anti, molina2015information, molina2016entanglement}.
On the other hand, the cMPS provides a variational class of non-Gaussian wave functional which is  just adapted to the non-relativistic interacting QFTs in $ 1+1$ dimensions.
In the case of relativistic QFTs, the cMPS construction suffers from regularization ambiguity. One can still use cMPS to study the low energy limit of the theory in practice by introducing a UV cut-off \cite{haegeman2010applying,stojevic2015conformal}. But still by construction, using the cMPS approach, one can not capture the short-distance behavior of the system.
Moreover, defining a UV cut-off by itself is in contrast with the purpose of working directly in the continuum.

In this paper, motivated by \cite{tilloy2021relativistic}, we study the one-parameter family of cMPS generated by ER which maps a free non-relativistic theory at IR scale to a free relativistic theory at UV. We will find that at the UV scale, the resulting wave functional is exactly the variational ansatz known as \emph{relativistic} cMPS (RCMPS) introduced in \cite{tilloy2021relativistic} which is adapted to relativistic QFTs in $ 1+1$ dimensions.

\subsection{Entanglement Renormalization in Continuum}

%$ Entanglement ~ Renormalization ~ in~ Continuum .$---
cMERA \cite{haegeman2011} was originally introduced as an ansatz
wave functional for the ground states of QFT Hamiltonians.
The same as the ER that corresponds to MERA tensor network, the continuous version of it implements  a real-space RG in the continuum. MERA on a lattice can also be visualized as a quantum circuit \cite{evenbly2009algorithms}. In this representation, the physical state can be obtained by evolving a simple product state with no entanglement that factories with respect to the lattice sites -usually considered as "all sites $0$"- by a unitary operator to create entanglement at different scales. 
The generalization to the continuum is conceptually straightforward. 
To describe cMERA first assume a QFT and impose a UV cut-off $ \Lambda$. 
%We are always working in the Hilbert space defined by the fields in the theory with the UV cut-off.
It is required to start with a finite  $ \Lambda$ in order to define the process but, in the end, it can be sent to infinity.
One parameter family of scale-dependent states is produced through continuous unitary evolution in scale $u$ as
\begin{equation}\label{4}
    \ket{\Psi (u)} = U (u, u_{IR}) \ket{\Omega}=\mathcal{P} e^ { -i \int _{u_{IR}} ^ u (K(s) + L ) ~ds} \ket{ \Omega}
\end{equation}
where the symbol $ \mathcal{P}$ is path ordering and $ \ket{\Omega}$ is the IR state that is the continuum limit of a product state on the lattice that contains no entanglement between spatial regions, and the UV state is what describes the system we are studying, usually, the ground state of the system. Moreover, it has been shown that any spacetime symmetry of the ground state is also a symmetry of the cMERA representation of it \cite{hu2017spacetime}.
Only the difference between UV and IR limits is fixed as 
$ u_{UV} - u_{IR} = O(\log \xi\Lambda )$ when $\xi$ is the correlation length of the theory.
It is convenient to set  $ u_{UV}= 0$ and $ u_{IR} = - O(\log \xi\Lambda )$ . For critical systems $ u_{IR} \rightarrow - \infty$.

On the other hand, $L$ is the generator of the scale transformation in spacial directions %(non-relativistic dilation operator)
and $ K(u)$ is the so-called entangler (or disentangler, depending on the direction of the RG flow) which  contains the variational parameters of the cMERA. 
The IR states is scale-invariant, thus $ L \ket{\Omega} =0$. 
Consider a set of field operators of the theory $ \psi (x)$, $ \psi^\dagger (x)$ satisfying 
$ [\psi (x) , \psi^\dagger (y)]_{\pm} = \delta (x-y)$ with $ + (-)$ for fermions (bosons). If the IR state is the vacuum of this set of annihilation and creation operators $ i.e. ~\psi (x) \ket{\Omega} = 0$ for all $ x$, 
the generator of scale transformation can be read as
\begin{equation}\label{2}
    L = - \frac{i}{2} \int \psi ^\dagger (x) x \frac{d \psi (x)}{dx} - x \frac{d \psi ^\dagger (x)}{dx} \psi (x) ~dx.
\end{equation}
Although some steps have been taken towards finding the form of the entangler operator for interacting theories, both at the perturbative level \cite{cotler2016gaussian, cotler2019entanglement, cotler2019renormalization} and non-perturbatively \cite{Fernandez-Melgarejo:2020fzw}, it has only been explicitly studied for free theories 
%of bosonic, fermionic and gauge fields 
\cite{haegeman2011, Franco-Rubio:2019nne}. The entangler operator for quadratic interactions is the generator of Bogoliubov transformation given by 
\begin{equation}\label{3}
      K(u) =  \frac{i}{2}  \int  dk~ \big(g(k, u)  \psi ^\dagger _k\psi ^\dagger _{-k}
      - g ^*(k, u) \psi_{-k} \psi_{k} \big)
\end{equation}
where $ \psi_k = \frac{1}{\sqrt{2 \pi}} \int  dx~ e^{-ikx} \psi (x)  $
 %is the Fourier transformation of $ \psi(x)$ 
 and 
 $ g(k/\Lambda, u) $ is even and odd in its first argument for bosons  and fermions, respectively. Finally, we mention that the cMERA unitary process provides a RG flow for the operators as
 \begin{equation}\label{6}
     \frac{d O(u)}{du} = - i [K(u) + L ~, O(u)]
 \end{equation}
when the physical or bare operators of the theory are defined at the UV scale.

\subsection{ Continuous MPS}
% $Continuous~Matrix ~ Product~ States.$---
The cMPS was originally proposed in \cite{verstraete2010continuous} by Verstraete and Cirac as a variational ansatz for the ground state of non-relativistic QFT Hamiltonians in $1+1$ dimensions. It can be obtained as the continuum limit of a certain family of MPS which is selected in such a way to have a valid continuum limit.

The most generic form of a MPS for a lattice with $N$ sites is given by 
\begin{equation}
    \ket{\psi} = \sum _{i_1 ,..., i_N} Tr[ A_1 ^{i_1} A_2 ^{i_2} ... A_N ^{i_N}] \ket{ i_1 ,i_2, ... , i_N}
\end{equation}
where $ A_n ^{i_n}$ are $ D\times D$ complex matrices containing the variational parameters of this ansatz.
%The matrix $B$ is the boundary operator that encodes the boundary conditions (B.C.). For the system with periodic B.C., $ B = I_D$ and with open B.C., it is given in terms of two arbitrary boundary vectors at $ x= \pm L/2$ as $ B = \ket{v_R} \bra{v_L}$.
Therefore, the MPS representation of the many-body wave function is specified by just $ O(D^2)$ variational parameters instead of exponential growth with $N$ which makes it a powerful variational ansatz for interacting theories. 
To find a generalization of MPS in the continuum,
we can approximate the QFT on a line of length $L$ by a lattice with lattice spacing $ \epsilon$ and $ N = L/\epsilon$ sites
\footnote{At each site of the lattice, there is a bosonic (or fermionic) mode $ a_i$ obeys $ [a_i , a_j^\dagger]_\pm = \delta_{ij}$ and thus, the Hilbert space spanned by $ \{ \ket{n_i}\}$ where $ \ket{n_i}$ corresponding to having $ n_i$ particles on that site. The many-body state $ \ket{ i_1 ,i_2, ... , i_N} $ can be rewritten as $ a_1 ^{i_1}  a_2 ^{i_2} ... a_N ^{i_N} \ket{\boldsymbol{0}}$, where $ \ket{\boldsymbol{0}} = \otimes _ {n =1} ^N \ket{0}_n $.}.
On this lattice, one can define a family of MPS as 
$ A^0  _i = I + \epsilon Q( i\epsilon)$ and 
$A^n _i = \frac{1}{n!} \big( \sqrt{\epsilon} R( i \epsilon)\big)^n $for $n \geq 1$.
By taking the $ \epsilon \rightarrow 0$ limit of it, one can find the class of cMPS as
\begin{equation}\label{5}
    \begin{split}
 \ket{ \psi [ Q, R]} = Tr_{aux} \big\{ \mathcal{P} \exp \int _{-L/2} ^ {L/2}  dx &
        \\
\big( Q(x) \otimes I +R(x)& \otimes \psi ^\dagger (x)\big) \big\} \ket{\Omega}      
    \end{split}
\end{equation}
where $ Tr_{aux}$ denotes a partial trace over the auxiliary system where the matrices $ Q$ and $R$ act. For the translational invariant cMPS the matrices $Q,~ R$ are position independent. 
The field $ \psi(x)$ is the continuum limit of the rescale modes 
%$ \psi (i \epsilon) = a_i / \sqrt{\epsilon}$, that
satisfying $ [\psi (x), \psi ^ \dagger (y)]_\pm = \delta (x-y)$, and $ \ket{\Omega}$, the empty vacuum 
%is the continuum limit of $ \ket{\boldsymbol{0}}$ that 
defined as $ \psi(x) \ket{\Omega} =0$ for all $x$, the same as the IR state of the cMERA. 
The expectation value of local operators and in particular the Hamiltonian on the cMPS representation of the ground state can be easily expressed in terms of the matrices $Q$ and $R$. 
In particular, all normal ordered correlation functions of local field operators can be deduced from a generating functional as
$ \langle : F[ \psi^\dagger (x), \psi (y)]: \rangle = F\big[ \frac{\delta}{\delta \bar{j} (x)}, \frac{\delta}{\delta j (y)} \big] \mathcal{Z}_{ \bar{j}, j} \big|_{\bar{j}, j =0}$,
when its explicit form can be given in terms of the cMPS matrices
\begin{equation}\label{7}
    \mathcal{Z}_{ \bar{j}, j} = Tr \Big\{ \mathcal{P} \exp \big[\int dx~  T+ j(x)~ R\otimes I + \bar{j}(x)~ I \otimes \bar{R}\big] \Big\}
\end{equation}
where $ T = Q \otimes I + I \otimes \bar{Q} + R \otimes \bar{R}$ is the cMPS transfer matrix \cite{haegeman2013calculus}.
To find the cMPS approximation of the ground state, it is just needed to minimize the expectation value of the Hamiltonian over the cMPS matrices. After that, correlation functions can be straightforwardly computed
\footnote{The same as MPS, the cMPS representation has gauge freedom that one can use to impose certain conditions on the cMPS matrices, including symmetry conditions. Moreover, for the continuum version, the left orthogonality condition of MPS can be read as 
$ Q(x) + Q^\dagger (x) + R^\dagger (x) R(x) =0 $ for all $x$.}.
By increasing $ D$, one can find a better approximation of the ground state.
In the last decade, several optimization algorithms have been developed to study a number of theories,  both bosonic and fermionic 
\cite{haegeman2010applying, draxler2013particles, quijandria2014continuous, chung2015matrix, quijandria2015continuous, haegeman2017quantum, rincon2015lieb, chung2017multiple, draxler2017continuous, ganahl2017continuous, ganahl2017continuous2, ganahl2018continuous, tuybens2022variational2}.
As mentioned, the cMPS provides an efficient variational ansatz for non-relativistic QFTs. It is not adapted to relativistic theories because of a lack of sensitivity to short-distance behavior.  One can look at \cite{tilloy2021variational,tilloy2021relativistic} for a complete explanation of the difficulties of applying the cMPS to the relativistic QFTs.

\begin{figure}
\centering
%\begin{tikzpicture}[line width=.6pt,line cap = round,scale=1.5,transform shape]

\begin{tikzpicture}[>=stealth, line width=.6pt,line cap = round,scale=.75,transform shape]
\draw[very thick,NordCyan ] (-.4,0) -- (6.6,0);
\draw[thick,very thick,NordYellow ] (-.4,6) -- (6.6,6);
\draw (0.1,0.1) .. controls (1.8,3) .. (2,5.9);
\draw (5.9,0.1) .. controls (4.2,3) .. (4,5.9);

\draw[->] (-.5,5.5) -- (-.5,3);
\node at (-.8, 3.3) {$u$};
\node at (-.9, 6) {$u_{IR}$};
\node at (-.9,0 ) {$u_{UV}$};
\shadedraw[left color= NordYellow, right color= NordBrightCyan, draw=white](0.1,0.1) .. controls (1.8,3) .. (2,5.9) --  (3.,5.9) -- (3,0.1) -- (0.1,0.1);
\shadedraw[left color= NordBrightCyan, right color= NordYellow ,  draw=white]
(3,0.1) --(3,5.9) --  (4,5.9) .. controls (4.2,3).. (5.9,0.1) -- (3,0.1);

%\filldraw[fill= NordCyan, draw= NordCyan](2.5,0.15) .. controls (2.7,3) .. (2.8,5.85) --  (3.2,5.85) .. controls (3.3,3).. (3.5,0.15) -- (2.5,0.15);
\draw[NordBrightCyan, very thick] (3,0.14)--(3,5.86);

\node at (3, 6.3) {$ cMPS [Q,R]$};
\node at (3, -.3) {$ relativistic~cMPS [\Tilde{Q},\Tilde{R}]$};

\node at (3, 1.8)  {\resizebox{.32\hsize}{!}{$ e^{- i \int du  (K(u)+L)}$}};
\end{tikzpicture}
\caption{Entanglement renormalization group flow of the class of cMPS}
\end{figure}
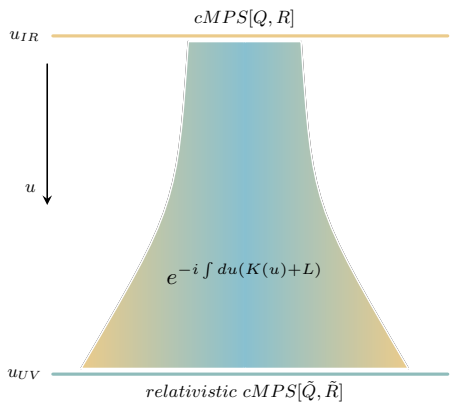

\subsection{cMERA RG flow of the class of  cMPS}

One can look at the cMPS representation of the ground state of a non-relativistic QFT as a non-Gaussian state generated by transforming the ground state of the free part of the Hamiltonian. This way of interpreting the cMPS is proposing to have a modification of the cMPS as a representation of the ground state of a relativistic QFT by transforming the ground state of the free relativistic QFT. There is a  cMERA RG flow that relates the ground states of the non-relativistic and relativistic free theories to each other.   In the following, we study the one-parameter family of the cMPS evolves with the corresponding cMERA evolution.
The cMERA formalism was originally formulated for infinite systems \cite{haegeman2011}. However,
its generalization to systems with open boundary conditions \cite{franco2021entanglement} and on a finite circle \cite{hung2021continuous} has been introduced more recently.
First, we work in the thermodynamic limit, $i.e.$, $ L \rightarrow \infty$, and after that, we will discuss the extension for the theories defined on a finite circle.

To proceed, we should find the generator of the RG flow in this case. Consider the free scalar field in the $ 1+1$ dimension. The Hamiltonian is given by 
\begin{equation}\label{8}
    H_{f.b.} = \frac{1}{2} \int dx~ \big[ \pi^2 (x) + (\partial_x \phi(x))^2 + m^2 \phi(x)^2 \big]
\end{equation}
where the field operator and its conjugate momentum satisfy $ [\phi (x), \pi (y)] = i \delta (x-y) $. One can expand them in terms of creation and annihilation operators $  a_k$ and $ a_k^\dagger$ satisfying $ [a_k, a^\dagger_{k'}] = 2\pi \delta(k-k')$.
The ground state of the theory is known to be the Fock space vacuum denoted by $ \ket{\boldsymbol{0}}_a$. 
In order to specify the cMERA representation of the ground state, we need to first define an unentangled reference state $ \ket{\Omega}$ in this theory. 
In general, one can define a Gaussian factorized state with width $ \Delta^{-1}$ as $ \psi (x) \ket{\Omega} =0 $ for all $x$, while
\begin{equation}\label{1}
    \psi (x) = \sqrt{\frac{\Delta}{2}} ~\phi (x) + i \sqrt{\frac{1}{2 \Delta}} ~\pi (x).
\end{equation}
By substituting \eqref{1} into \eqref{2} and \eqref{3},  we have the form of the cMERA Hamiltonian in terms of $\phi$ and $ \pi$.  The function $ g(k,u)$ in \eqref{3} is assumed to be real-valued in the form 
$ g(k,u) = \chi (u) \Theta (1- |k|/\Lambda) $, where $ \Theta (x)$ is the step function. 
By considering 
$ \ket{\psi (u=0) } = \ket{\boldsymbol{0}}_a$ in \eqref{4}, we find an ansatz to represent the exact ground state of the theory as a circuit cMERA. As the last step, one should apply the variational principle and minimize the energy
$ E = \bra{\psi (u=0)} H_{f.b.} \ket{\psi (u=0)}$ to exactly find $ \Delta$ and $ \chi(u)$.

In order to do the calculation, it is useful to go to the interaction picture where $L$ can be understood as the free part of the cMERA Hamiltonian while $ K(u)$ is the interacting part. One can rewrite the unitary evolution in scale in the  interaction picture as 
\begin{multline}
    U(u_1,u_2) = e^{-i u_1 L} ~\hat{U} (u_1, u_2) ~e^{ i u_2 L}
    \\
  = e^{-i u_1 L} ~ \mathcal{P} e^ { -i \int _{u_{2}} ^ {u_1} \hat{K}(u)~du}  ~e^{ i u_2 L}
\end{multline}
where
$ \hat{K} (u) = e^{i u L} K(u) e^{ -i u L}$ can be read of as
$\hat{K} (u) = \frac{i}{2} \int dk~ g( k e ^{-u}, u) \big ( a_k ^\dagger a^ \dagger _{-k} - a_{-k} a_{k}\big )$.
Finally, by requiring $ \delta E / \delta \chi (u) =0$ for every $ u$, we find that 
$ \Delta = \sqrt{ \Lambda ^2 + m^2}$ and 
$ \chi (u) = \Lambda ^2 e^{2u} / 2 ( \Lambda ^2 e^{2u} 
 + m^2)$.
 
Before going ahead to find the RG flow of the class of cMPS, in order to find the renormalized operators via the evolution in scale, it is good to know that
\begin{equation}\label{12}
    e^{-iuL} \psi (k) e^{iuL} = e^{-u/2} \psi (k e^{-u})
\end{equation}
and under the action of $ \hat{U}  (u, u_{IR})$
\begin{equation}\label{13}
     \begin{pmatrix}
  a_k\\ 
 a_{-k} ^ \dagger
\end{pmatrix}  \longrightarrow \begin{pmatrix}
  \cosh \theta (u) & - \sinh \theta (u)\\ 
 - \sinh \theta (u)& \cosh \theta (u)
\end{pmatrix} \begin{pmatrix}
  a_k\\ 
 a_{-k} ^ \dagger
\end{pmatrix} 
\end{equation}
where 
$ \theta(u)= \int _{u_{IR}} ^ u ds~ g (ke^{-s} , s)$ and 
$\theta (u=0) = \ln \sqrt{ \frac{\Delta}{ \omega_k}} $, while $ \omega_k = \sqrt{k^2 + m^2}$.

Now, we are ready to define a one-parameter family of states by relating them to the IR state 
%$ \ket{\Psi (u_{IR})}= \ket{\psi[Q,R]}$
through the entangling evolution in scale as 
$ \ket{\Psi (u)}=  U(u, u_{IR})\ket{\psi[Q,R]}$. 
One can explicitly expand the path order in \eqref{5} and  obtain 
\begin{equation}
\begin{split}
\ket{\psi[Q,R]} = \sum_{n=0}^\infty \frac{1}{n!} \int _{-\infty} ^\infty 
   dx_1 dx_2... & dx_n
    \\
    \Phi_n (x_1, x_2,..., x_n) \psi^\dagger (x_1)  \psi^\dagger &(x_2) ...\psi^\dagger(x_n) \ket{\Omega}
\end{split}
\end{equation}
while
$ \Phi_n (x_1,..., x_n) = Tr [ \mathcal{P} \{ e^{\int _{-\infty} ^\infty  Q(y) dy} R(x_1) ... R(x_n)\}] $. 
To find the explicit form of $ \ket{ \Psi (u)}$, it is enough to determine the transformation of the 
$ \psi^\dagger (x_1) ...\psi^\dagger(x_n) \ket{\Omega} $ 
under the action of the unitary evolution which is 
$ \psi^\dagger (x_1,u) ...\psi^\dagger(x_n,u) \ket{\psi (u)} $, where we define
$\psi^\dagger (x,u) = U( u, u_{IR}) \psi^\dagger (x) U ^{-1}( u, u_{IR}) $ and 
$ \ket{\psi (u)} = U( u, u_{IR}) \ket{\Omega}$.
In particular by using 
\eqref{12} and \eqref{13},
one can obtain that  at the UV scale 
$ \psi (x,0) = e^{u_{IR}/2} a(x e^{u_{IR}}) $ where $ a(x) = 1/\sqrt{2\pi}\int dk e^{ikx} a_k$ is defined to be the Fourier transform of the annihilation operator $ a_k$. By construction, we also have
$ \ket{\psi(0)} = \ket{\boldsymbol{0}} _a$.
In the end, we obtain the UV state as 
\begin{equation}
\begin{split}
 \ket{ \Psi [\tilde{Q}, \tilde{R}]} = Tr_{aux} \Big\{\mathcal{P} \exp \int _{-\infty} ^ {\infty}  &dx
    \\
    \big( \tilde{Q}(x) \otimes I  +\tilde{R}&(x) \otimes a ^\dagger (x)\big)\Big\} \ket{\boldsymbol{0}}_a 
\end{split}
\end{equation}
while
\begin{equation}\label{11}
  \tilde{Q}(x) = e^{- u_{IR}} Q (x e^{-u_{IR}}) ~~~  \tilde{R}(x) =  e^{- u_{IR} /2} R (x e^{-u_{IR}}) . 
\end{equation}
It is nothing but the class of  RCMPS introduced in \cite{tilloy2021relativistic} as an ansatz to approximate the ground state of a relativistic QFT without requiring any additional  UV cut-off, and thus, the result is valid even at high momenta. As the operator $ a(x)$ has the same algebra as $ \psi(x)$, RCMPS inherits the properties of the class of cMPS by replacing $ \psi (x)$ with $ a(x)$. 
Specifically, the correlation function of the $ a(x), a^\dagger(x)$ can be obtained via the same generation functional \eqref{7}. 
The only important point is that since $ a(x)$ is not local in terms of $ \phi$ and $ \pi$, 
the computation of the expectation value of the Hamiltonian is more difficult than in the non-relativistic cases.
Moreover, the naive optimization, which works well for the standard cMPS, fails for RCMPS and one should use some more precise methods like the tangent space approach \cite{vanderstraeten2019tangent}. In \cite{tilloy2021relativistic}, RCMPS was used to study the self-interacting $ \phi ^4$ theory and provided some remarkable  results.

Finally, one can also check that by defining 
$ H(u=0) = H_{f.b.}$, we will find at IR scale the Hamiltonian of the non-relativistic free boson as
$ H(u_{IR}) = \frac{1}{2 \tilde{m}} \int dx~ \partial_x \psi ^\dagger (x) \partial _x \psi (x) + \mu \int dx ~\psi ^\dagger (x) \psi (x)$, while $\tilde{m} = m e^{2 u_{IR}}$ and $ \mu = m$ is the so-called chemical potential.

\emph{RCMPS for Fermionic Theories.---}
%$RCMPS~ for ~ fermionic~ theories.$---
The free relativistic fermions in the $ 1+1$ dimensions given by Dirac Hamiltonian
\begin{equation}
    H_{Dirac} = \int dx ~ [ \bar{\psi} (x) \sigma_2 \partial_x \psi (x) + m \bar{\psi} \psi]
\end{equation}
where $ \psi = (\psi_1, \psi_2)^T$ is the two-component complex fermions and $ \bar{\psi} = \psi ^ \dagger \sigma_3$.
Here, one can choose the unentangled state as $ \psi_1(x) \ket{\Omega} = 0 =\psi_2^\dagger(x) \ket{\Omega} $. The standard class of cMPS is defined as 
\begin{equation}\label{9}
    \begin{split}
\ket{\psi[Q, R_1,R_2]} = 
    Tr_{aux} \big\{\mathcal{P} 
    \exp \int dx~ \big( & Q(x) \otimes I
        \\
      +R_1(x) \otimes \psi_1 ^\dagger (x)+ R_2(x) &\otimes \psi_2(x)\big) \big\} \ket{\Omega}.  
    \end{split}
\end{equation}
To find the related class of states appropriate for relativistic theories, we need to find the exact form of the RG flow such that $ \ket{\boldsymbol{0}} = U(u=0, u_{IR}) \ket{\Omega}$, where $ \ket{\boldsymbol{0}}  $ is the exact ground state of the Dirac Hamiltnian.
The entangler is given as 
$ K(u) = i \int dk g(k,u)\big( \psi_1^\dagger \psi_2(k) + \psi_1(k) \psi_2(k)\big) $. In this case, the Bogoliubov angle is antisymmetric and we can suppose its form as 
$ g(k,u) = k \chi(u) \theta (1- |k|/\Lambda)$. The same as a free boson, one can find $ \chi(u)$ by minimizing the expectation value of the Hamiltonian \cite{haegeman2011}.
Moreover, one can derive that 
$  e^{-iuL} \psi_{1,2} (k) e^{iuL} = e^{-u/2} \psi_{1,2} (k e^{-u}) $
while $ \psi _i(k)$ is the Fourier transform of $ \psi _i(x)$ ,
and under the action of the unitary evolution in the interaction picture
\begin{equation}
     \begin{pmatrix}
  \psi_1(k)\\ 
   \psi_2(k)
\end{pmatrix}  \longrightarrow \begin{pmatrix}
  \cos \theta_f (u) & - \sin \theta_f (u)\\ 
  \sin \theta_f(u)& \cos\theta_f (u)
\end{pmatrix} \begin{pmatrix}
 \psi_1(k)\\ 
   \psi_2(k)
\end{pmatrix} 
\end{equation}
where
$ \theta_f(u)= \int _{u_{IR}} ^ u ds g (ke^{-s} , s)$ and 
$\theta_f (u=0) = \frac{1}{2} \arcsin (-k/ \omega_k)$. 
By considering \eqref{9} as the IR state, we can find the fermionic RCMPS at UV scale, $ i.e., u=0$ as 
\begin{equation}
    \begin{split}
        \ket{\Psi[\tilde{Q}, \tilde{R}_1,\tilde{R}_2]} =
    Tr_{aux} \big\{\mathcal{P} 
    \exp\int dx~ 
    \big( &\tilde{Q}(x) \otimes I 
    \\
    +\tilde{R}_1(x) \otimes b_1 ^\dagger (x)+ \tilde{R}_2&(x) \otimes b_2(x)\big) \big\} \ket{\boldsymbol{0}}.  
    \end{split}
\end{equation}
while $ \tilde{Q}$ and $ \tilde{R}$ defined by \eqref{11}, and $ b_{1,2}(x)$ are the Fourier transform of the $ b_{1,2}(k)$ which can be found in terms of $  \psi_{1,2}(k)$ as 
$ b_{1}(k) = \alpha_k \psi_1(k) + \beta _k \psi_2(k)$
and $ b_{2}(k) = -\beta_k \psi_1(k) + \alpha _k \psi_2(k)$ while
$ \alpha_k= -k / \sqrt{k^2 + (\omega_k-m)^2}$
$ \beta_k= (m-\omega_k) / \sqrt{k^2 + (\omega_k-m)^2}$. One can check that 
$ [H, b_1^\dagger(k)] = \omega_k b_1^\dagger(k)$ and 
$[H, b_2(k)] = \omega_k b_2(k)$ or in other words, the set of operators $ b_{1,2}(k)$ are the modes diagonalizing the Dirac Hamiltonian.

\emph{ RCMPS on a Circle.---}
% $RCMPS~ on~a~ circle.$---
Finding the RCMPS on a circle requires having the cMERA RG flow for  relativistic free fields on the circle.
In \cite{hung2021continuous}, it has been shown that if a Gaussian cMERA describes the ground state of a theory on a line, the ground state of the same theory on a circle has a cMERA representation as well. Furthermore, the cMERA entangler can be obtained using the method of images. 
The unentangled reference state defined as $ \psi (x) \ket{\Omega^c} =0$ for $ x \in [0, l_c)$ when $ \psi(x)$ is again given by \eqref{1}.
The entangler has the form of 
$ K^c(u)= \frac{i}{2}\sum _{n \in \mathbb{N}} \tilde{g}^c(n,u)
[\psi_n^\dagger \psi_{-n} ^\dagger - \psi_n\psi_{-n}]$ where
$ \psi_n = 1/\sqrt{l_c} \int_0 ^ {l_c} dx e^{-ik_n x} \psi (x)$ for $ n \in \mathbb{Z}$ and $ k_n = 2 \pi n / l_c$.
The entangling profile on the circle  is defined as
$ \tilde{g}^c(x,u) = 1/ \sqrt{l_c} \sum _n e^{ik_n x} \tilde{g}^c (n,u)$  can be obtained from the one on the line $ g(x,u)$ through the method of images as 
$ \tilde{g}^c(x,u) = \sum _{n\in \mathbb{Z}} g(x+n l_c,u) $. It implies that
$ \tilde{g}^c(n,u) = g(k,u) \big|_{k=k_n}$.
Following the procedure described above, one can generalize RCMPS to  an ansatz as a variational class to approximate the ground state of the relativistic theory on a finite circle as
\begin{equation}
    \ket{ \Psi [Q,R]}^c = Tr_{aux} \big\{\mathcal{P} e^{ \int _{0} ^ {l_c} dx~ \big( \tilde{Q}(x) \otimes I +\tilde{R}(x) \otimes a ^{c\dagger} (x)\big)} \big\} \ket{\boldsymbol{0^c}}_a 
\end{equation}
where 
$ a^c(x) \ket{\boldsymbol{0^c}}_a  =0$ for all $ x \in [0,l_c)$ and 
$ a^c(x)$ is defined as the Fourier transform of the modes which diagonalize the free theory on the circle \cite{hung2021continuous}.

\emph{Discussion.--}
In this paper, we could obtain the class of RCMPS via a RG flow generated by an appropriate cMERA circuit. 
They can be used to approximate the ground states of the relativistic QFTs in $1+1$ dimensions containing both bosonic theories like the sine-Gordon model and fermionic ones such as the Gross-Neveu and Thirring models. 
Moreover, since the Gaussian cMERA is known in higher dimensions
for all bosonic, fermionic, and gauge fields \cite{haegeman2011, Franco-Rubio:2019nne}, the procedure above can provide a way to find appropriate wave functionals for relativistic theories in higher dimensions, especially, the relativistic version of the continuous PEPS in $ 2+1$ dimensions.
Furthermore, an alternative approach to RCMPS  for relativistic theories is the interacting cMERA (icMERA) \cite{Fernandez-Melgarejo:2020fzw}. 
It can be found by modifying the entangler and going beyond the Bogoliubov transformation by adding the terms  generate n-tuplet transformation in fields. Thus, the icMERA evolution is the combination of two Gaussian and non-Gaussian unitaries, exactly the same as RCMPS. Although for icMERA, the important point is the fact that to date, we do not know for a given theory, up to what n-tuplet interacting terms are exactly needed to capture the full non-perturbative structure of the theory. But in the case of RCMPS, the form of the ansatz is fixed for all the families of the relevant theories.
On the other hand, there is freedom in choosing the entangling profile of the entangler operator of the cMERA. In particular, there is a specific choice that leads to another class of states called magic cMERA \cite{ zou2019magic} which is already shown that has the same UV structure as the standard cMPS. Moreover, its entangler by itself has the continuous matrix product operator representation.
Therefore, studying the connection between them might even help us for a better understanding of the interacting disentangler.
In the end, we would like to point out that since cMERA is connected to AdS/CFT, it would be desirable to study the possible gravity dual of the states of the form of RCMPS.

\emph{Acknowledgements.--}
I  would like to thank Guglielmo Lami, Nishan Ranabhat, Mandana Safari, and Emanuele Tirrito for useful discussions.
I would also like to thank Kyriakos Papadodimas for his support and  CERN-TH for hospitality during the first stage of this work.

\bibliographystyle{apsrev4-1} % Tell bibtex which bibliography style to use
\bibliography{references.bib} % Tell bibtex which .bib file to use (this one is some example file in TexLive's file tree)

\end{document}